\documentclass[preprint,superscriptaddress,aps,amssym]{revtex4}

\usepackage{latexsym}
\usepackage{textcomp}
\usepackage{graphicx,wrapfig}
\usepackage{epsfig}
\usepackage{dcolumn}
\usepackage{amsmath}

\usepackage{longtable}
\usepackage[usenames]{color}
\usepackage{amsfonts}
\usepackage{amssymb}

\usepackage{natbib}
\setcitestyle{numbers,sort&compress}

\usepackage[normalem]{ulem}
\usepackage{hyperref}
\usepackage{epstopdf}
\usepackage{float}
\usepackage{natbib}
\usepackage{multirow}
\usepackage{verbatim}

\input colordvi

\begin{document}

\title{Strong chemisorption of CO$_2$ on B$_{10}$-B$_{13}$ planar-type clusters}
\author{Alexandra B. Santos-Putungan}
\email[Author to whom correspondence should be addressed. Electronic address:] {absantos1@up.edu.ph}
\affiliation{Materials Science and Engineering Program, University of the Philippines Diliman, Quezon City, Philippines} 
\affiliation{Institute of Mathematical Sciences and Physics, University of the Philippines Los Ba\~{n}os, College, Los Ba\~{n}os, Laguna, Philippines}

\author{Nata\v{s}a Stoji\'c}
\author{Nadia Binggeli}
\affiliation{The Abdus Salam International Centre for Theoretical Physics, Trieste, Italy}

\author{Francis N.~C.~Paraan}
\affiliation{National Institute of Physics, University of the Philippines Diliman, Quezon City, Philippines}

\begin{abstract}
  {An {\it ab initio} density functional study was performed investigating the adsorption of CO$_2$ on the neutral boron B$_{n}$ ($n = 10-13$) clusters, characterized by planar and quasiplanar ground-state atomic structures. For all four clusters, we found strong chemisorption energy of CO$_2$ reaching 1.6~eV for B$_{12}$ at the cluster edge sites with the adsorbed molecule in the plane of the cluster. A configuration with chemisorbed dissociated CO$_2$ molecule also exists for B$_{11}$ and B$_{13}$ clusters. The strong adsorption is due to the bending of the CO$_2$ molecule, which provides energetically accessible fully in-plane frontier molecular orbitals matching the edge states of the clusters. At the same time, the intrinsic dipole moment of a bent CO$_2$ molecule facilitates the transfer of excess electronic charge from the cluster edges to the molecule.
}
\end{abstract}

\startpage{1}
\endpage{ }
\maketitle

\section{Introduction}\label{intro}

Carbon dioxide (CO$_2$) is a greenhouse gas that contributes greatly to global warming. As the use of carbon-based fuel is a primary source of energy, it is desirable to develop technologies for efficient capture and sequestration of CO$_2$ produced from such sources. Significant efforts have been carried out to study adsorption of CO$_2$ on different materials including complicated structures such as covalent organic frameworks \cite{Zeng2016, Lan2010} and metal organic frameworks \cite{Zhang2016, Saha2010}. In this respect,  CO$_2$ adsorption on boron clusters and surfaces offers an interesting alternative \cite{Sun2014PCCP, Sun2014JPCC} which deserves further investigation.  

Boron, like its neighboring element carbon, possesses a remarkable variety of structures that could be of use in a wide range of applications \cite{Zhang2012, Endo2001, Carter2014}. Bulk  boron polymorphs are mainly composed of 3D-icosahedral B$_{12}$ cage structures as basic building blocks \cite{Bullett1982, Perkins1996}, while small boron clusters prefer planar-type aromatic/antiaromatic structures \cite{Zhai2003, Sergeeva2014}. 

In fact, neutral and charged clusters B$_{n}^{(+,-)}$, with ${n \leq 15}$, have been predicted theoretically \cite{Boustani1997, Ricca1996, Kiran2005,Tai2010}, and confirmed experimentally (or by combined experimental and theoretical studies) \cite{Zhai2003, Oger2007, Tai2010, Romanescu2012}, to be planar or quasiplanar. For ${{n} > 15}$, competing low-energy isomers start to occur, in particular for the positively charged clusters B$_{16}^+$ to B$_{25}^+$ which were reported to have ring-type structures, based on mobility measurements \cite{Oger2007}. On the other hand, the negatively charged B$_{n}^-$ clusters have shown to systematically conserve planar-like structures up to at least ${{n}=25}$ by joint photoelectron spectroscopy and quantum chemistry calculations \cite{SerAveZha11,PiaLiRom12, PopPiaLi13,PiaPopLi14}. Moreover, the neutral clusters B$_{16}$ and B$_{17}$ clusters are found to display planar-type geometries based on vibrational spectroscopy studies \cite{Romanescu2012}; in this case, the smallest 3D-like (tubular) structure was suggested to occur for B$_{20}$ \cite{Kiran2005PNAS}. Recently, B$_{27}^-$, B$_{30}^-$, B$_{35}$, B$_{35}^-$, B$_{36}$ and B$_{36}^-$ clusters have been discovered to possess quasiplanar geometries through combined experimental and theoretical studies \cite{PiaHiLi14, Li2014JACS, Li2014Ange, Li2015JCP}, while the B$_{40}$ cluster has been observed to occur with a fullerene structure \cite{Zhai2014}. Such quasiplanar clusters can be viewed as embryos for the formation of 2D boron sheets (borophenes) \cite{PiaHiLi14, Li2014JACS}. Several borophene polymorphs and boron nanotubes have been theoretically predicted \cite{Yang2008, Quandt2005, XWu2012, Weir1992} and also experimentally grown \cite{Ciuparu2004, Weir1992, Patel2015, Mannix2015}.

Previous computational studies have revealed an interestingly strong CO$_2$ adsorption behavior on some theoretical models of surfaces of solid $\alpha$-B$_{12}$ and $\gamma$-B$_{28}$ \cite{Sun2014PCCP} and relatively strong CO$_2$ binding energies on B$_{40}$ and B$_{80}$ fullerenes \cite{Dong2015, Gao2015, Sun2014JPCC}. For the most common boron planar type of clusters, as well as for 2D-boron sheets, however, chemical binding of CO$_2$ was theoretically predicted so far only in the case of chemically engineered systems, namely for charged transition metal (TM)-atom centered boron-ring clusters, TM\textendash B$_{8-9}^-$ \cite{Wang2015}, and for Ca-, Sc- coated boron sheets \cite{Tai2013}.

In the current work, we show the existence of strong chemical binding of the CO$_2$ molecule to the aromatic/antiaromatic planar-type B$_{n}$ clusters (${n}=10~\textrm{to}~13$). By means of first-principle calculations and by varying the CO$_2$ initial position, we identify various chemisorbed and physisorbed configurations. We find that the strong chemisorption occurs for all four clusters when the adsorbed CO$_2$ molecule is in the plane of the cluster, close to its edge, and that the strongest adsorption energy reaches 1.6~eV in the case of B$_{12}$. For B$_{11}$ and B$_{13}$ adsorption with dissociated CO$_2$ is also found to occur at some edge sites. We rationalize the mechanism of the strong adsorption as due to the strong and matching planar character of frontier orbitals of both the cluster and bent CO$_2$ molecule, together with the favorable redistribution of electronic charge in excess at the edges of the cluster, in the presence of the dipole moment of the bent CO$_2$.

\section{Methodology and systems}\label{method}

\subsection{Computational details}

All calculations were carried out using first-principles plane-wave pseudopotential density functional theory (DFT) method, as implemented in the Quantum ESPRESSO package \cite{Giannozzi2009}. The spin-polarized Perdew-Burke-Ernzerhof (PBE) \cite{Perdew1996} exchange-correlation functional within the generalized gradient approximation (GGA) was employed. We used scalar-relativistic  Vanderbilt ultrasoft pseudopotentials \cite{Vanderbilt1990} generated from the following atomic configurations: $2s^{2}2p^{1}$ for B, $2s^{2}2p^{2}$ for C and  $2s^{2}2p^{4}$ for O. A non-linear core correction was included in the B pseudopotential. We employed a cubic supercell with sides of 21~\AA\ for all calculations to avoid cluster interactions. A 1~$\times$~1~$\times$~1 Monkhorst-Pack \textbf{k}-point mesh was used with a Gaussian level smearing of 0.001 Ry. Threshold for electronic convergence was set to 10$^{-7}$~Ry, and structures were optimized until the forces on each atom were below 10$^{-4}$~Ry/a.u.

The CO$_{2}$ adsorption energy ($E_\textrm{ads}$) on the B clusters was computed as \cite{Sun2014JPCC}:
\begin{equation} \label{eq:E_ads}
E_\textrm{ads}=E_{\textrm{B}_{n}-\textrm{CO$_{2}$}}-{E}_{\textrm{B}_{n}}-{E}_\textrm{CO$_{2}$},
\end{equation}

\noindent where $E_{\textrm{B}_n-\textrm{CO}_2}$ is the total energy of the atomically relaxed system consisting of the B$_{n}$ cluster and adsorbed CO$_{2}$ molecule, $E_{\textrm{B}_n}$ is the total energy of the isolated (relaxed) B$_{n}$ cluster, and $E_{\textrm{CO}_{2}}$ is the total energy of the CO$_2$ molecule in the gas phase. Convergence tests for the plane-wave expansion of the electronic orbitals indicated that changing the kinetic energy cut-off from 64~Ry to 96~Ry resulted in $E_\textrm{ads}$ changes within 1~meV. We used the former wave-function cut-off, together with a 384-Ry cut-off for the augmentation charge density, in all calculations reported here.

\subsection{Geometry and relative stability of the B$_\textrm{10-13}$ clusters}

\begin{figure}[!h]
\centering\includegraphics[width=0.7\linewidth]{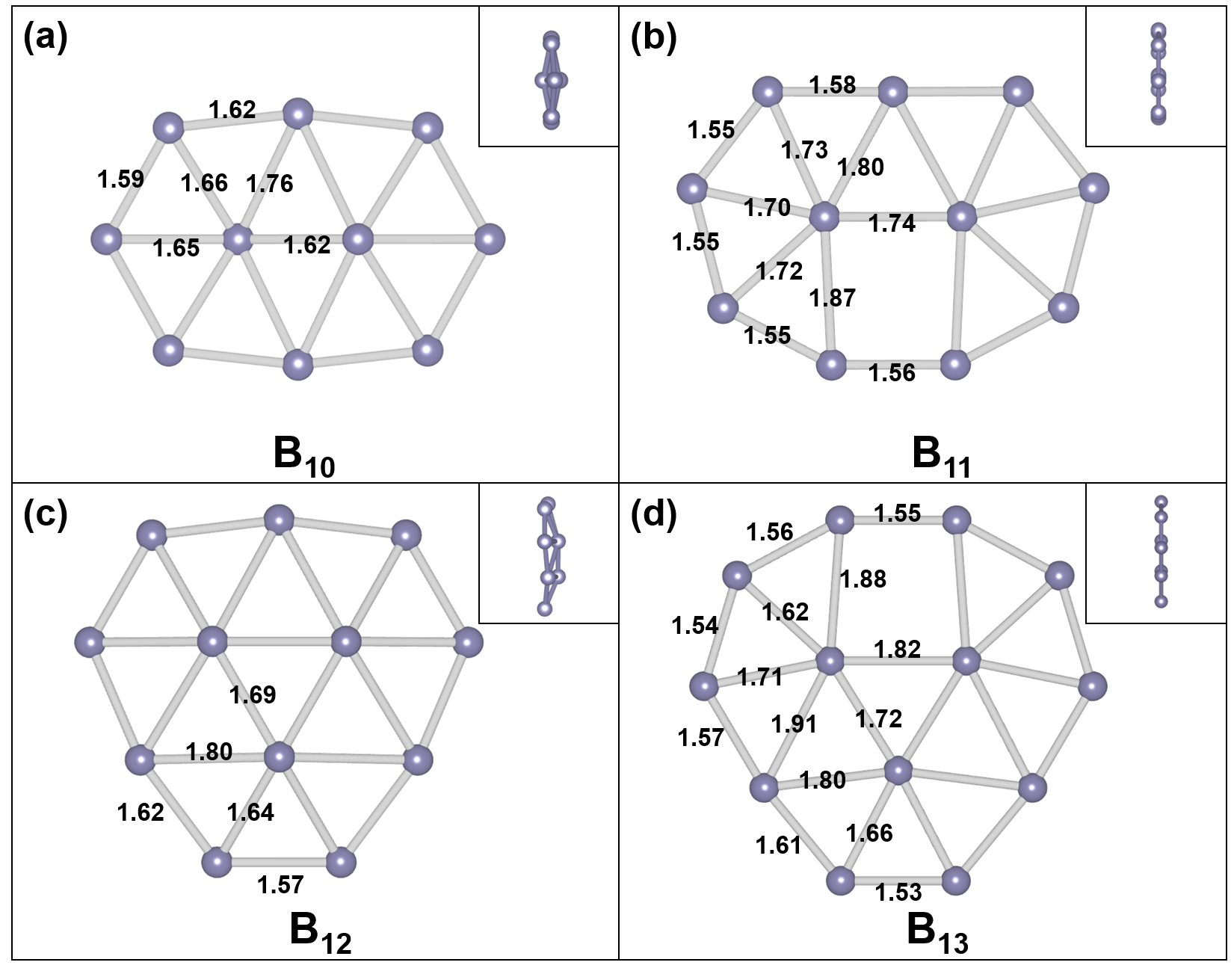}
\caption{Obtained optimized structures of B$_{n}$ clusters (${n}=10-13$). Specific B\textendash B bond lengths, in \AA, are also indicated for each cluster. Insets show the side view of the cluster, demonstrating that B$_{11}$ and B$_{13}$ clusters exhibit planar structures, while B$_{10}$ and B$_{12}$ are quasiplanar with some of the atoms displaced by 0.31 and 0.34~\AA\ from the cluster plane for B$_{10}$ and B$_{12}$ clusters, respectively.}
\label{fig:bondlengths}
\end{figure}

The initial boron-cluster structural configurations were constructed based on previous work \cite{Tai2010} that catalogued the stable structures of B$_{n}$ clusters (for ${n \leq 13}$). We performed structural optimization resulting in the lowest-energy cluster geometries and bond lengths, shown in  Fig.~\ref{fig:bondlengths} that are consistent with the results in Ref.~\cite{Tai2010}. It can be seen that B$_{10}$ and B$_{12}$ clusters exhibit quasiplanar structures, while B$_{11}$ and B$_{13}$ clusters have planar structural geometries. Moreover, B$_{12}$ and B$_{13}$ clusters are characterized by three inner atoms that are compactly bound forming an inner triangle. The longest B\textendash B bonds of $\geq$1.8~\AA\ existing in these clusters belong to B$_{11}$ and B$_{13}$ clusters, and form a square configuration within the cluster (see Fig.~\ref{fig:bondlengths}). Among the B$_n$ clusters studies, B$_{12}$ is the energetically most stable with a binding energy of 5.37~eV/atom (calculated binding energies are given in Supplementary material, Part I: Table~S1). 

\section{Results and discussions}\label{results}

\subsection{Chemisorption of CO$_2$ on B$_{n}$ clusters}

We considered different initial configurations of CO$_2$ relative to the B$_{n}$ clusters including various adsorption sites and orientations of the molecule. We found strong chemisorption of the CO$_2$ molecule along the contour edges of the B$_{n}$ clusters. In addition, physisorption states of the CO$_2$ molecule were observed at larger distances from the cluster or when placed on top of the B-cluster plane. The strong binding of CO$_2$, with adsorption energies between $-1.6$ and $-1$~eV, is a common feature for all four B$_{n}$-clusters.

\begin{figure}[!h]
\centering\includegraphics[width=0.7\linewidth]{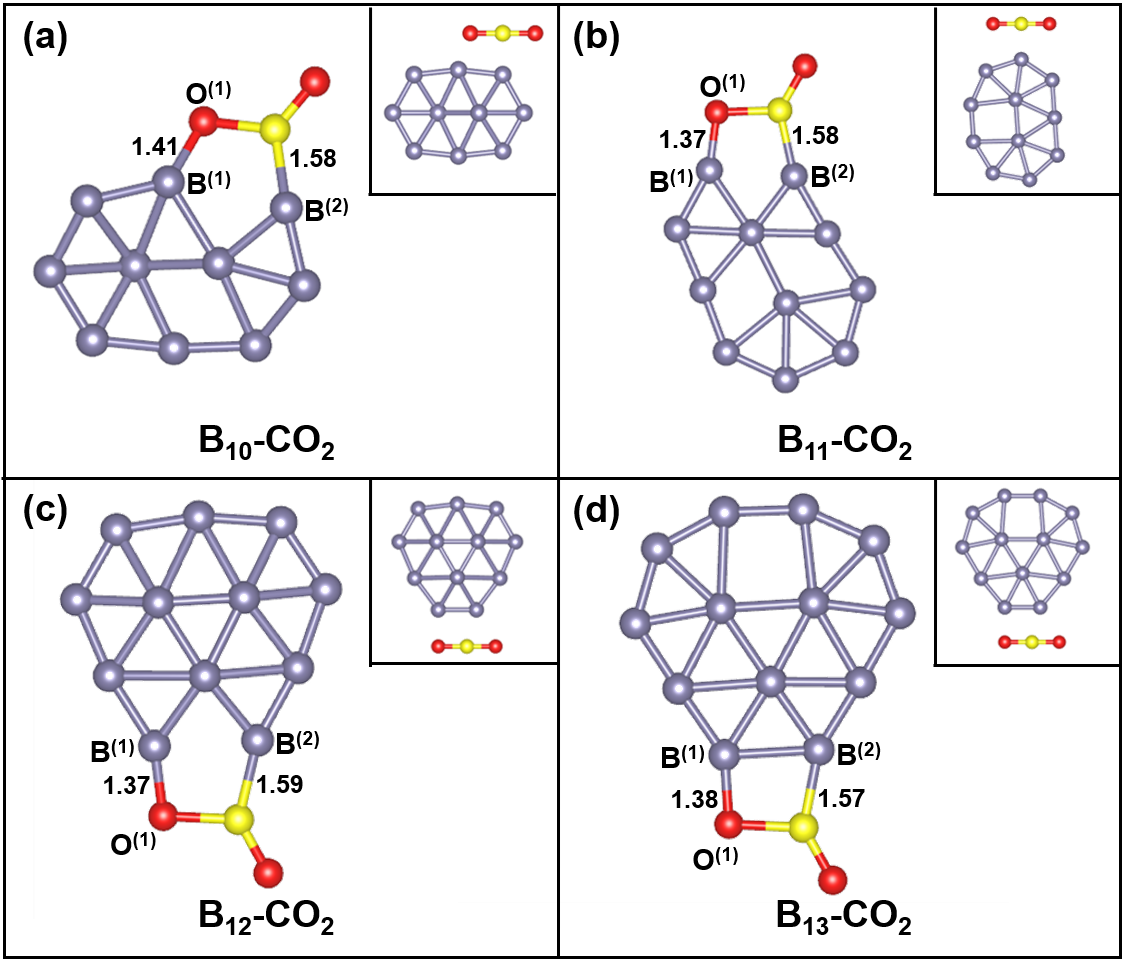}
\caption{Obtained optimized structures of CO$_2$ with B$_{n}$ clusters for the strongest adsorption, where B, C and O atoms are shown in grey, yellow and red, respectively. The  distances between the cluster and molecule are given in angstroms. Insets represent initial positions prior to the interaction of the CO$_2$ molecule with B clusters, with the molecule placed in the cluster plane at less than 2~\AA\ distance from the cluster. Boron bonds shorter than 2~\AA\ are represented by rods.}
\label{fig:init_final}
\end{figure}

Figure~\ref{fig:init_final} shows the obtained optimized configurations of the B$_{n}$\textendash CO$_{2}$ systems characterized by the strongest adsorption energy ($E_\textrm{ads}$, shown in Table~\ref{tab:E_ads}), together with their corresponding initial configurations, shown as insets. The strongest adsorption overall was found for the B$_{12}$\textendash CO$_{2}$ system with a chemisorption energy of $-1.60$~eV, followed by close values of about $-1.4$~eV for B$_{11}$ and B$_{13}$, and somewhat weaker, but still robust chemisorption on B$_{10}$. Besides similar strong values of the CO$_{2}$ adsorption, all four B$_{n}$\textendash CO$_{2}$ systems share common features regarding the adsorption geometry. Thus, chemisorption occurs when the  CO$_2$ molecule is initially placed near the edge sites and in-plane with respect to the B$_{n}$ cluster (see the insets in Fig.~\ref{fig:init_final}) for all clusters. Furthermore, final configurations indicate that chemisorbed B$_{n}$\textendash CO$_2$ systems tend to keep a planar geometry. The CO$_2$ molecule bends by a similar angle of $\sim$122\textdegree~for all B clusters considered as it chemisorbs on the B cluster. It should be noted that this angle corresponds to the equilibrium geometry predicted theoretically for the negatively charged CO$_2$ molecule \cite{GutBarCom98}. Following the formation of a C\textendash B and O\textendash B bond (with lengths of $\sim$1.6 and $\sim$1.4~\AA, respectively), the O\textendash C bond lengths of the molecule (initially 1.18~\AA) also elongate asymmetrically to $\sim$1.2~\AA\ (the  O\textendash C bond further away from the cluster) and to $\sim$1.5~\AA\ (for the O\textendash C bond that is linked to the B cluster). Distances between B atoms at which O and C atoms are bound (denoted B$^{(1)}$ and B$^{(2)}$ respectively, in Fig.~\ref{fig:init_final}, with the binding O denoted O$^{(1)}$) increase by $0.3~-~0.7$~\AA\ with respect to their bond lengths in isolated clusters. Other edge chemisorption sites were also found for all the four clusters (with $E_\textrm{ads}$~$<$~$-1.10$~eV).

\begin{table}[!h]
\centering
\caption{Strongest adsorption energies (in eV) obtained for the relaxed configurations with adsorbed CO$_2$ molecule on the B$_{n}$, ${n}=10-13$, clusters and with dissociated molecule (CO + O) in the cases of B$_{11}$ and B$_{13}$ (second line). The adsorption energies correspond to the final configurations shown in Figs. \ref{fig:init_final} and \ref{fig:dissoc}. The adsorption energy of the dissociated CO$_2$, $E_\textrm{ads}^\textrm{dissociated}$, was obtained using Eq.~\ref{eq:E_ads}.}

\begin{tabular}{l  c  c  c  c}
\hline
     & \textbf{B$_\textrm{10} $} & \textbf{B$_\textrm{11} $} & \textbf{B$_\textrm{12} $} & \textbf{B$_\textrm{13} $}\\
\hline
$E_\textrm{ads}$ (eV) & $-1.11$ & $-1.42$ & $-1.60$ & $-1.43$\\
\hline
$E_\textrm{ads}^\textrm{dissociated}$ (eV) & --- & $-2.19$ & --- & $-1.66$\\
\hline
\end{tabular}
\label{tab:E_ads}
\end{table}

\begin{figure}[!h]
\centering\includegraphics[width=0.7\linewidth]{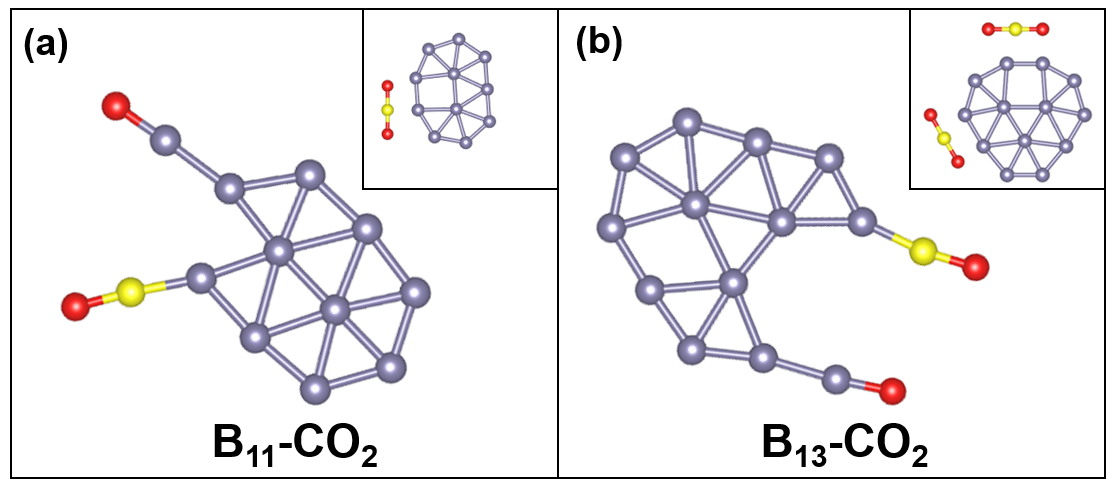}
\caption{Obtained optimized structures of the CO$_2$ molecule adsorbing on (a) B$_{11}$ and (b) B$_{13}$ clusters where dissociation of the molecule occurs. Insets show the initial position prior to the interaction with the molecule placed in the cluster plane  at a  distance of less than 2~\AA\ from the cluster.}
\label{fig:dissoc}
\end{figure}

Dissociation of CO$_2$ was also observed in B$_{11}$ and B$_{13}$ clusters at some specific B sites, wherein some of B bonds broke in order for the dissociated O and C\textendash O fragments to bind to the (deformed) cluster, as shown in Fig.~\ref{fig:dissoc}. For B$_{11}$ and B$_{13}$ clusters with dissociated CO$_2$, the chemisorption energies ($E_\textrm{ads}^\textrm{dissociated}$) are $-2.19$~eV and $-1.66$~eV, respectively. 

\begin{figure}[!h]
\centering\includegraphics[width=0.4\linewidth]{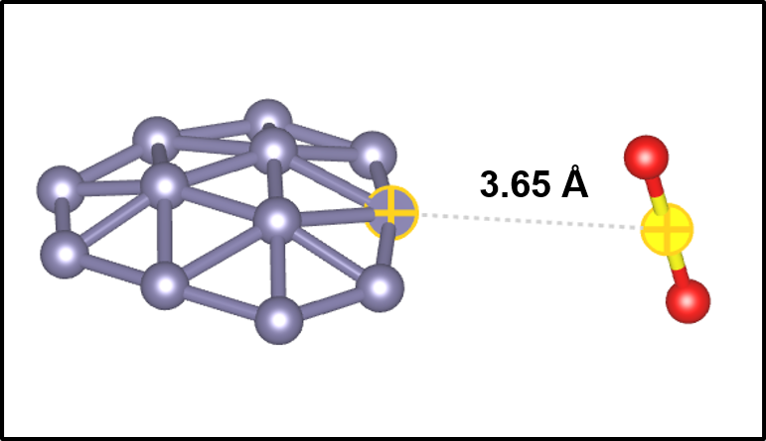}
\caption{Representative image of a typical physisorption state of CO$_2$ molecule on B$_{12}$ cluster obtained when the molecule is initially placed near an edge atom of the cluster, and rotated 90\textdegree ~out of the cluster plane. The CO$_2$ molecule maintains its linear structure as it moves away from the cluster.}
\label{fig:physi_correct}
\end{figure}

We also found physisorbed CO$_2$ configurations with physisorption energies ranging from $-11$ to $-30$~meV for distances between 3.5 and 4~\AA~from the B$_{n}$ cluster (measured as the smallest interatomic separation). The physisorption configurations include the CO$_2$ molecule placed above the cluster or placing the C of the molecule further away in the cluster plane, with the O atoms in or out of the cluster plane (as shown in Fig.~\ref{fig:physi_correct} for the case of B$_{12}$). An example describing the in-plane physisorption and chemisorption states of CO$_2$ on B$_{12}$ cluster is given in Part II of Supplementary material.

The binding energies we have found here for the chemisorbed CO$_2$ molecule on the neutral, metal-free planar-type clusters (in the range 1.1 \textendash~1.6 eV for B$_{10-13}$) are significantly larger than previously obtained for 3D-type cluster structures ($\sim$0.4~eV for B$_{40}$ and $\sim$0.8~eV for B$_{80}$ \cite{Sun2014JPCC, Dong2015, Gao2015}). To the best of our knowledge, this is the first study that provides evidence of the strong chemical binding of the CO$_2$ molecule to planar-type B clusters, although good adsorption was theoretically found for a few diatomic molecules on selected B clusters \cite{ValFarTab15, SunWanLi13, SloKanPan10}. The CO$_2$ binding energies to B$_{11-13}$ we obtained are also larger than those reported for the chemically engineered TM\textendash B$_\textrm{8-9}^-$ clusters and for the metallated/charged fullerenes ($1.1-1.2$~eV) \cite{Wang2015,Dong2015, Gao2015}. We note that previous studies have indicated that charging the Boron fullerenes or engineered TM\textendash B$_\textrm{8-9}$ negatively tends to enhance the adsorption of CO$_2$ \cite{Gao2015, Wang2015}, which suggests that even stronger adsorption could be obtained for B$_{n}^-$ planar clusters.

Furthermore, we expect the strong bonding character obtained here for CO$_2$ to B$_\textrm{10-13}$ to persist for the larger planar-type B clusters. In fact, we have examined the binding properties of CO$_2$ to a semi-infinite Boron $\alpha$-sheet \cite{TanIsm07,note_sheet} and also found chemisorption with ${E_\textrm{ads}\approx-0.3}~\textrm{eV}$ and a similar type of CO$_2$ adsorption geometry (including the $\sim$122\textdegree~O$^{(1)}$\textendash C\textendash O bond angle) at the edge of the Boron sheet \cite{note_sheet}. The latter may be viewed as the edge of a planar B$_{N}$ cluster in the limit of large $N$.

Finally, we stress that the large chemisorption energy we find is a robust feature of the system that persists even in the presence of Hubbard on-site interactions that are implemented via GGA + U calculations \cite{note_U}. The interactions provided by U increase the CO$_2$ HOMO-LUMO gap (next section), and are actually found to enhance the adsorption strength (binding energy) of the CO$_2$ molecule to the B clusters.

\subsection{Electronic properties of the distorted and undistorted isolated systems}

In order to better understand the strong chemisorption of CO$_2$ on all considered B planar-type clusters, we have examined the atomic-orbital-resolved density of states of the isolated clusters and bent molecule, focusing on the atoms participating in the formation of the chemisorption bonds. As we have seen in the previous section, the CO$_2$ bond angle changes from 180\textdegree~(free molecule) to approximately 122\textdegree~in the chemisorbed geometry, which is indicative of a negative charging of the molecule. Moreover, the bending itself of the CO$_2$ molecule significantly modifies its electronic spectrum, and in particular considerably reduces its HOMO-LUMO gap \cite{Tai2013b}. In fact, an important point to note is that, when the molecule bends, the previously degenerate (from linear CO$_2$) highest-occupied and lowest-unoccupied $\pi$ states of the molecule both split into in-plane and out-of-plane orbitals, leaving exclusively O and C $2p$-related in-plane molecular orbitals as the frontier orbitals of the 122\textdegree - bent molecule (see Supplementary material, Fig.~S2 and also Fig.~\ref{fig:pdos_mo}(a)).

The splitting, after the molecule bends, of the lowest-unoccupied $\pi$ ($p_{z}$,$p_{y}$) level, in particular, is very large (3.7~eV) compared to the HOMO level splitting (0.4~eV, Fig.~S2) and the overall HOMO-LUMO gap also drastically decreases (by 6.6~eV in our calculations, Fig.~S2(b) and Fig.~\ref{fig:pdos_mo}(a)) with respect to the linear molecule (Fig.~S2(a)). Figure~\ref{fig:pdos_mo}(a) shows, for the resulting bent CO$_2$, the in-plane components of the $2p$-related O$^{(1)}$ and C projected density of states (PDOS) along the B\textendash C bond direction ($p_{y}$ component) and perpendicular to it ($p_{x}$ component). The corresponding molecular orbitals for the levels closest to the gap are also displayed in the figure. As can be seen, the bent CO$_2$ molecule has fully planar-type HOMO and LUMO states (denoted as H$_\textrm{A1}$ and L$_\textrm{A1}$ in Fig.~\ref{fig:pdos_mo}), in strong contrast with the linear CO$_2$ molecule (Fig.~S2(a)). The PDOS in Fig.~\ref{fig:pdos_mo}(a) also shows that, while the HOMO of the bent molecule retains very strong O$^{(1)}$ and C $2p_{y}$-orbital character, the LUMO exhibits both a strong $2p_{y}$ component and a substantial $2p_{x}$ component (both antibonding) from the  O$^{(1)}$ and C atoms.

In Fig.~\ref{fig:pdos_mo}(b), we display, for the isolated B$_{12}$ cluster, the same type of $p_{x}$ and $p_{y}$ in-plane components of the density of states projected on the $2p$-orbitals of B$^{(1)}$ and B$^{(2)}$ atoms (the $2p_{z}$ component is shown in the Supplementary material, Fig.~S3). Such in-plane states are the ones which may interact/hybridize with the frontier orbitals of the bent CO$_2$. In Fig.~\ref{fig:pdos_mo}(b), we also display for the levels closest to the HOMO-LUMO gap and having the highest in-plane PDOS, the corresponding molecular orbitals. These states are characterized by lobes protruding over the cluster's edge within the cluster plane.

It can be observed from Fig.~\ref{fig:pdos_mo}(b) (and comparison with the full $p$-state PDOS in Fig.~S3(b)) that there is an especially large density of in-plane orbitals of the {\it{peripheral B atoms}} (B$^\textrm{(1)}$ and B$^\textrm{(2)}$) in the upper (2 to 3~eV) region of the cluster occupied-state spectrum. We note that previous calculations indicated that the B clusters which we are considering have in total in the occupied spectrum only 3 to 4 $p_{z}$-type (out-of-plane) molecular orbitals \cite{Zubarev2007}, delocalized over all cluster atoms, which is also what we find. The high density of in-plane $p_{x}$ and $p_{y}$ orbitals from peripheral (B$^\textrm{(1)}$ and B$^\textrm{(2)}$) atoms in the top (2 to 3~eV) part of the cluster occupied-state spectrum is a feature common to all four clusters considered in this work.

\begin{figure}[!h]
\centering\includegraphics[width=0.35\linewidth]{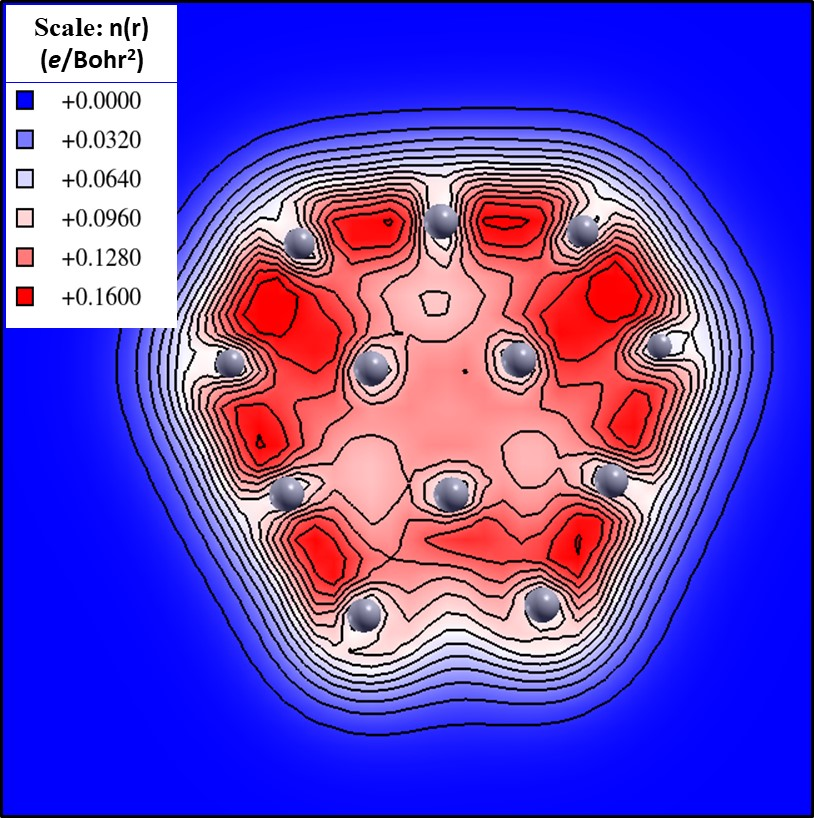}
\caption{Electronic charge density contour plot calculated for an isolated B$_{12}$ cluster with the same distorted atomic structure as in the adsorbed B$_{12}$\textendash CO$_2$ system. The distortion is occuring mostly at the atoms which take part in the binding (the bottom two B atoms in the plot).  It can be seen that the electronic charge density is systematically largest at the cluster contour edges leaving thus an extended positively charged area in the central part of the cluster. One can also observe that the adsorption of the CO$_2$ molecule causes it to lose its 3-fold symmetry.}
\label{fig:B12distortedchargedens}
\end{figure}

The in-plane molecular states of the cluster in the energy region [$-5$~eV, $-1$~eV], in Fig.~\ref{fig:pdos_mo}(b), strongly contribute to the electronic charge density of the cluster along its contour edge. In Fig.~\ref{fig:B12distortedchargedens}, we display the electronic charge density of the isolated B$_{12}$ cluster with the distorted geometry as in the adsorbed B$_{12}$\textendash CO$_2$ system. The electronic charge distribution is similar to that of the free/undistorted B$_{12}$ cluster (Fig.~S1 in Supplementary material); it is largely concentrated at the contour edges of the cluster. This inhomogeneous electronic distribution makes the contour edges negatively charged and leaves the inner B atoms with a reduced electron density. These properties are observed in all four clusters investigated here (Fig.~S1 in the Supplementary material).

\begin{figure}[H]
\centering\includegraphics[width=0.7\linewidth]{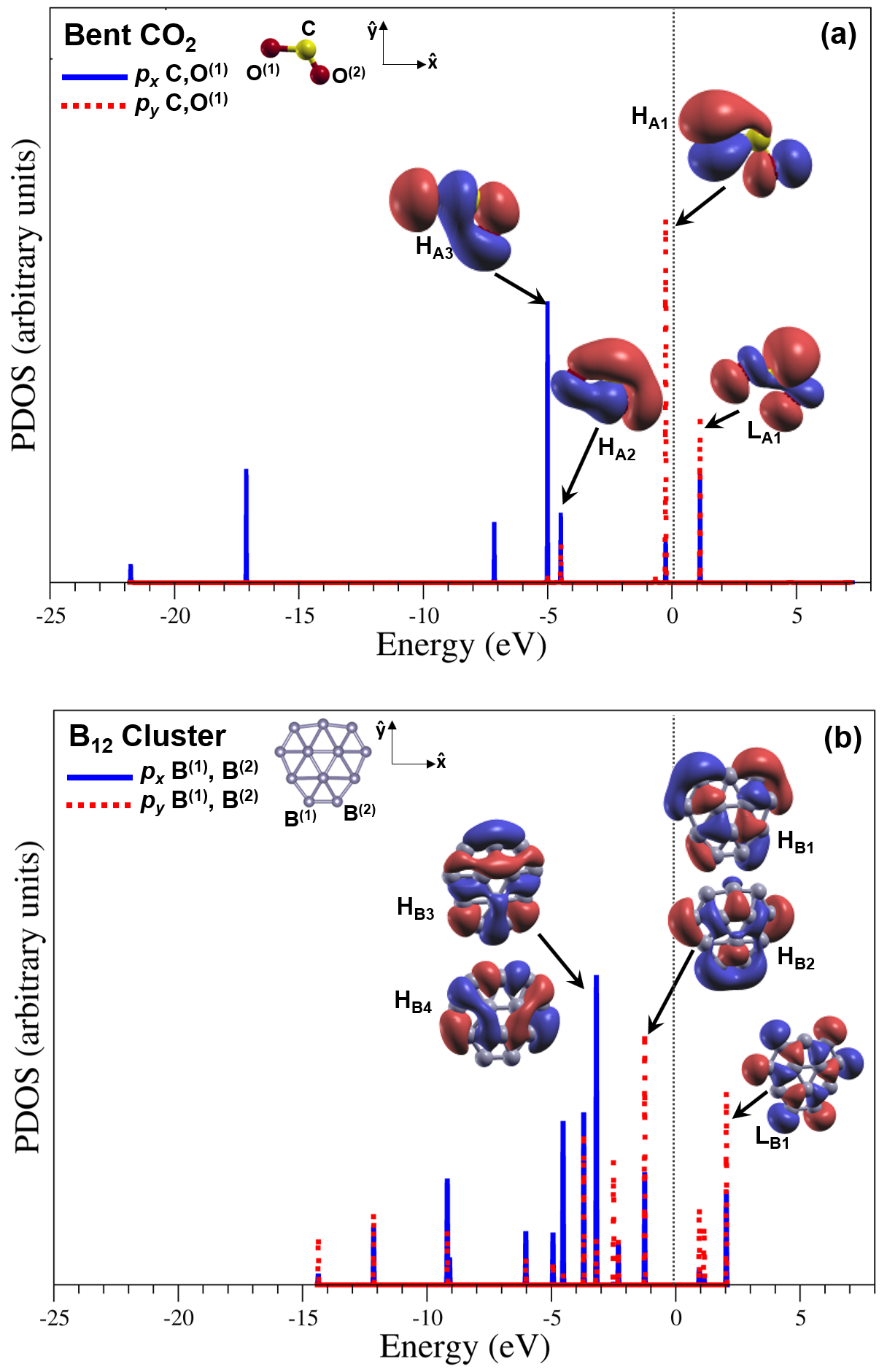}
\caption{Atomic $2p_{x}$ and $2p_{y}$ projected density of states (PDOS) of the isolated bent CO$_2$ molecule (a) and B$_{12}$ cluster (b) summed over the two atoms directly involved in the chemisorption bonds in the configuration shown in Fig.~\ref{fig:init_final}c, i.e., O$^{(1)}$ and C in panel (a) and B$^\textrm{(1)}$ and B$^\textrm{(2)}$ in panel (b). The bent molecule and cluster are also shown with the corresponding $\hat{{x}}$ and $\hat{{y}}$ directions: the $y$-direction is aligned with the B\textendash C bond and the $x$-axis is perpendicular to it, still remaining in the plane of the cluster. Some of the occupied, $E < 0$, and empty, $E > 0$, states (probability density with orbital phase change) of the bent CO$_2$ molecule and of the B$_{12}$ cluster are shown next to their respective PDOS. The isosurface level is set to 0.001 $e$~\AA$^{-3}$.}
\label{fig:pdos_mo}
\end{figure}

\subsection{Discussion of the chemisorption mechanism}

\begin{figure}[!h]
\centering\includegraphics[width=0.4\linewidth]{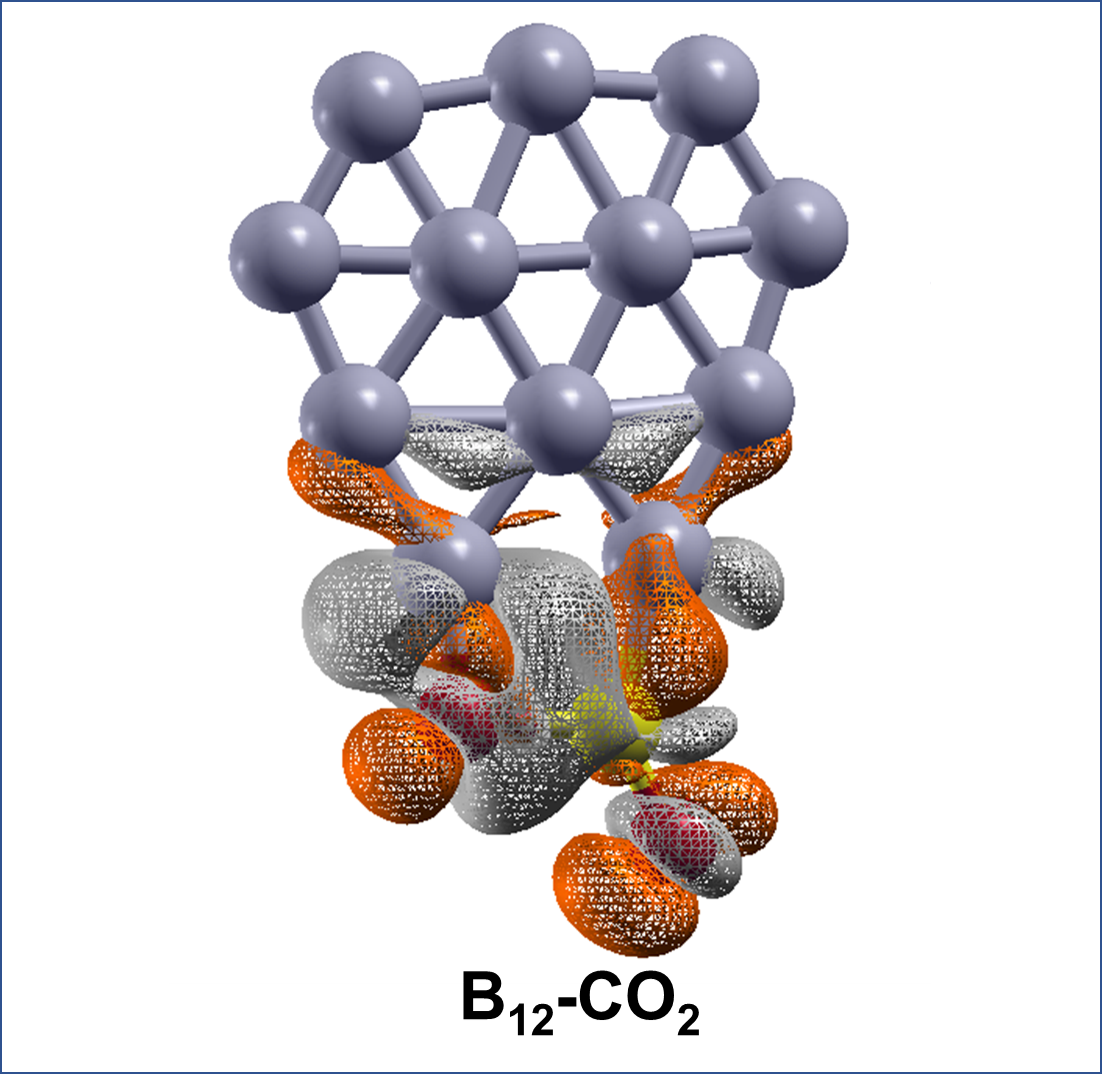}
\caption{Differential electron density isosurface ($\Delta$$\rho$) for the B$_{12}$\textendash CO$_2$ system (see text). Gray color represents electron deficient region ($\Delta$$\rho < 0$), while orange denotes electron rich region ($\Delta$$\rho > 0$) with respect to the isolated B$_{12}$ cluster and CO$_2$ molecule. A large electron rich region can be observed for the adsorbed CO$_2$ molecule, indicating that CO$_2$ acquired excess electrons becoming effectively negatively charged. The isosurface level is set to 0.004 $e$~\AA$^{-3}$. It can be observed that the overall shape of the electron-gain (orange) differential charge density isosurface in the region of the CO$_2$ molecule resembles that of probability density of the LUMO of bent CO$_2$ (refer to L$_\textrm{A1}$ of Fig.~\ref{fig:pdos_mo}).}
\label{fig:chargediff}
\end{figure}

To identify the dominant CO$_2$ molecular orbital involved in the chemisorption, we examined the differential charge density, i.e., the difference between the charge density of the chemisorbed B$_{n}$\textendash CO$_2$ system and that of the isolated B$_{n}$ and CO$_2$. In Fig.~\ref{fig:chargediff}, we present differential charge-density isosurfaces illustrating the electronic charge difference associated with the chemisorption of CO$_2$ on B$_{12}$. The shape of the energy-gain isosurface in the region of the CO$_2$ molecule has strong similarities with the probability density isosurface of the LUMO of the bent CO$_2$ molecule (refer to L$_\textrm{A1}$ of Fig.~\ref{fig:pdos_mo}). The LUMO CO$_2$ orbital will interact with some planar high-energy occupied molecular orbital(s) of the cluster (in Fig.~\ref{fig:pdos_mo}(b)) and, based on the probability densities of the molecular orbitals of the interacting B$_{12}$\textendash CO$_2$ system (the highest occupied states are shown in Fig.~S4 in the Supplementary material), we find that the L$_\textrm{A1}$ molecular orbital of CO$_2$ interacts (hybridizes) predominantly with the H$_\textrm{B3}$ molecular orbital of the cluster (see Fig.~\ref{fig:pdos_mo}(b)). These molecular orbitals have lobes protruding from the edges of the cluster/molecule with substantial orbital overlap suggesting that strong interaction between cluster and molecule can take place in this region.

From Fig.~\ref{fig:chargediff} it can be inferred that the  CO$_2$ molecule gained excess negative charge from the B cluster. We performed a Lowdin charge analysis, which (although it cannot provide exact values of the atomic charges in a hybridized system and is basis dependent) is useful to give charging trends. Thus, the C atom (binding with B$^\textrm{(2)}$) gained $0.27~e$, and the O atom (binding with B$^\textrm{(1)}$) gained $0.04~e$, while only a very small charge transfer for the other O atom is seen ($\sim$ 0.001~$e$). Similar total amount of Lowdin charge was lost by the B cluster. Strong charge transfer between the B structures and the chemisorbed CO$_2$ molecule has been reported earlier and related to the Lewis acid-base interactions \cite{Sun2014PCCP,Sun2014JPCC,Wang2015}. The electronic charge transfer from the edges of the cluster (with excess negative charge) to the molecule can be also rationalized considering that the bent CO$_2$, in difference to the linear molecule, has a net dipole moment which is substantial: 0.724~ea$_0$ \cite{MorHay79}. The positive end of the dipole is closer to the B cluster and the negative side further away from the cluster, facilitating the interaction with the edge sites of B cluster that exhibit higher electronic density.

In addition to the strong chemisorption of the full CO$_2$ molecule on B clusters, we also found cases where the molecule dissociated into C\textendash O and O fragments (Fig.~\ref{fig:dissoc}), each of which  is bound separately to the B cluster, having typical bond lengths of 1.2 and 1.5~\AA\ (for both B$_{11}$ and B$_{13}$), respectively. The dissociation is attributed to the presence of longer bond lengths and lower charge density of the clusters, together with the specific choice of adsorption sites closest to the long B\textendash B bonds. The dissociation of the molecule takes place at B\textendash B edges where the charge density of the cluster is relatively low (Fig.~S1 in the Supplementary material) and the B atoms have less bonding with other B atoms. Both B$_{11}$ and B$_{13}$ clusters have considerably smaller HOMO-LUMO gap values than the other two clusters which do not display dissociative adsorption (Table~S1 in Supplementary material). The smaller gap indicates higher chances of  interaction between the cluster and molecular states, allowing also  more varied types of adsorption configurations, as we observe in our calculations. 

\section{Conclusion}

We investigated the adsorption of CO$_2$ on B$_{n}$ ($n=10-13$) clusters by using first-principles density-functional theory. These clusters have been predicted theoretically and confirmed experimentally to have  planar or quasiplanar geometries. We obtained different chemisorbed and physisorbed configurations depending on the initial position of the CO$_2$ molecule. In particular, the chemisorption is obtained for an in-plane position of the molecule close to the cluster contour edges, with adsorption, thus, at the cluster edge sites. CO$_2$ chemisorbs strongly to all four clusters considered, while the strongest CO$_2$ binding energy, amounting to 1.6~eV, is calculated for B$_{12}$. The CO$_2$ chemisorption energies we found for the B$_{10-13}$ clusters are considerably larger than previously obtained for the neutral B$_{80}$ and B$_{40}$ fullerene-type clusters. To the best of our knowledge, this is the first time such strong chemical binding of CO$_2$ to the planar-type B clusters is evidenced. The CO$_2$ binding energies to B$_{11-13}$ we obtained are also larger than previously reported for the chemically engineered TM\textendash B$_{8-9}^-$ clusters and doped/charged B fullerenes. We explain the strong chemisorption by the planarity of the B clusters which are characterized by a high density of protruding occupied in-plane molecular-orbital states near the cluster gap, associated with peripheral B atoms, and excess electronic charge at the cluster edges. These properties facilitate binding with the bent CO$_2$ molecule, which has exclusively in-plane frontier orbitals and a non-vanishing dipole moment.

\section{Acknowledgements}\label{acknowledgements}
This work was funded by the UP System Enhanced Creative Work and Research Grant ECWRG 2018-1-009. A.B.S.-P. is grateful to the Abdus Salam International Centre for Theoretical Physics (ICTP) and the OPEC Fund for International Development (OFID) for the OFID-ICTP postgraduate fellowship under the  ICTP/IAEA Sandwich Training Educational Programme, and to the Philippines Commission on Higher Education (CHEd) for the Faculty Development Program (FacDev)\textendash Phase II.


\bibliographystyle{unsrtnat}

\begin{thebibliography}{53}
\bibitem[not()]{note_U}
{We included a Hubbard U calculation on the molecule localized $p$-related
  states ($U=U_{\textrm{O}-2p}=U_{\textrm{C}-2p}$) using a range of values
  (between 2 and 9~eV). This systematically increases the CO$_2$ LUMO-HOMO gap,
  and we found that the adsorbate binding energy also increases, giving, for
  example, for $U=3$~eV (and a HOMO-LUMO gap increase by 0.5~eV), a binding
  energy stronger by 0.2~eV for the CO$_2$ chemisorption on B$_{12}$}.
\end{thebibliography}

\end{document}